\documentclass[a4paper,11pt]{article}

\usepackage{pos}
\usepackage{graphicx} % Required for inserting images
\usepackage{amsmath}
\usepackage{caption}
\usepackage{subcaption}
\usepackage{bbm}
\usepackage{todonotes} % for boxes with todos
\usepackage[capitalise]{cleveref}
\usepackage{algorithm} % for pseudocode
\usepackage{algpseudocode} % for pseudocode
\usepackage{siunitx}

\crefname{section}{Sec.}{Secs.}

\newcommand{\nstep}{n_\text{step}}

\title{Toward Scalable Normalizing Flows for the Hubbard Model}
\ShortTitle{Toward Scalable Normalizing Flows for the Hubbard Model}

\author*[a,b]{Janik Kreit}
\author[a,b]{Andrea Bulgarelli}
\author[a,b]{Lena Funcke}
\author[b,c]{Thomas Luu}
\author[a,b]{Dominic Schuh}
\author[a,b]{Simran Singh}
\author[d]{Lorenzo Verzichelli}

\affiliation[a]{Transdisciplinary Research Area (TRA) Matter, University of Bonn, Germany}
\affiliation[b]{Helmholtz Institute for Radiation and Nuclear Physics (HISKP), University of Bonn, Germany}
\affiliation[c]{Institute for Advanced Simulation 4 (IAS-4), Forschungszentrum Jülich, Germany}
\affiliation[d]{Physics Department, University of Turin \& INFN - Turin unit, Italy}

\emailAdd{jkreit@uni-bonn.de}
\abstract{Normalizing flows have recently demonstrated the ability to learn the Boltzmann distribution of the Hubbard model, opening new avenues for generative modeling in condensed matter physics. In this work, we investigate the steps required to extend such simulations to larger lattice sizes and lower temperatures, with a focus on enhancing stability and efficiency. Additionally, we present the scaling behavior of stochastic normalizing flows and non-equilibrium Markov chain Monte Carlo methods for this fermionic system.}

\FullConference{%
The 42nd International Symposium on Lattice Field Theory,\\
2nd-8th November, 2025,\\
Tata Institute of Fundamental Research, Mumbai, India
}

\begin{document}
\maketitle

\section{Introduction}
\label{sec:introduction}
\noindent
The Hubbard model is a central framework in condensed-matter physics for describing strongly correlated electronic systems \cite{hubbard,Arovas_2022}. It can be formulated either in terms of a Hamiltonian or an action, with the latter typically referring to the finite-temperature auxiliary-field formulation. Despite its simple structure, numerical simulations in this formulation are challenging due to the multimodal structure of the Boltzmann distribution and the presence of large potential barriers. As a consequence, standard sampling methods such as Hybrid Monte Carlo (HMC) often suffer from ergodicity violations~\cite{wynen2019}, which can lead to biased estimates of physical observables.

Generative neural samplers (GNSs), including Normalizing Flows (NFs)~\cite{rezende2015variational,kobyzev2020normalizing} and autoregressive models~\cite{oord2016pixelrecurrentneuralnetworks}, have shown strong performance in representing Boltzmann distributions across a wide range of physical systems. Often referred to as \textit{Boltzmann generators}~\cite{doi:10.1126/science.aaw1147}, they have been applied in lattice quantum field theory~\cite{PhysRevD.100.034515,PhysRevLett.126.032001,cranmer2023advances}, statistical mechanics~\cite{PhysRevLett.122.080602,PhysRevE.101.023304}, string theory~\cite{Caselle:2023mvh, Caselle:2024ent}, and quantum chemistry~\cite{doi:10.1126/science.aaw1147,gebauer1,gebauer3}. Beyond accelerating sampling, GNSs enable direct estimates of thermodynamic observables~\cite{PhysRevLett.126.032001} and entanglement entropies~\cite{bialas2024r,Bulgarelli:2024yrz}. In parallel, non-equilibrium Markov chain Monte Carlo (NE-MCMC) methods~\cite{Caselle:2016wsw, Bonanno:2024udh, Bulgarelli_2025, Bonanno:2025pdp} remain effective in regimes where HMC suffers from ergodicity problems and strong autocorrelations~\cite{noneq_jerome}.

In this work, we investigate the performance of NE-MCMC methods, NFs, and SNFs for sampling the Boltzmann distribution of the Hubbard model. While NFs provide efficient generative proposals and NE-MCMC methods offer favorable scaling properties, SNFs combine both by augmenting deterministic flow transformations with stochastic updates \cite{Caselle:2022acb, Bulgarelli_2025}. This unified framework is particularly well suited for addressing ergodicity issues in Hubbard model simulations.

Building on earlier work on symmetry-enforced normalizing flows for the Hubbard model at small volumes~\cite{Schuh:2025gks,Kreit:2025cos}, we extend these studies to larger lattice volumes and lower temperatures by improving the numerical stability of the Hubbard action. We also present initial scaling results for NE-MCMC methods, NFs, and SNFs in this setting.

The remainder of these proceedings is organized as follows. In~\cref{sec:hubbard}, we review the Hubbard model. In~\cref{sec:qr}, we introduce an algorithm to improve numerical stability at low temperatures. The methods of NE-MCMC, NFs, and SNFs are discussed in~\cref{sec:method}, and their scaling results are presented in~\cref{sec:results}. Conclusions and an outlook are given in~\cref{sec:conclusion}.

\section{Hubbard Model}
\label{sec:hubbard}
\noindent
The Hubbard model is defined by the Hamiltonian
\begin{equation}
\label{eq:hamiltonian}
H = -\kappa \sum_{\langle x,y \rangle} \left( a_{x,\uparrow}^\dagger a_{y,\uparrow} + a^\dagger_{x,\downarrow} a_{y,\downarrow} \right)- \frac{U}{2} \sum_x \left(n_{x,\uparrow} - n_{x,\downarrow} \right)^2 \,,
\end{equation}
where $a^{\dagger}_{x,\sigma}$ ($a_{x,\sigma}$) denotes the fermionic creation (annihilation) operator for a particle with spin $\sigma \in \{ \uparrow, \downarrow \}$ at lattice site $x$, and $n_{x,\sigma} \equiv a^\dagger_{x,\sigma} a_{x,\sigma}$ is the corresponding number operator. The first term in~\cref{eq:hamiltonian} describes nearest-neighbor hopping with amplitude $\kappa$, while the second term represents an on-site interaction with coupling strength $U > 0$. In the following, we consider hexagonal, half-filled lattices which correspond to zero chemical potential. For all simulations, we set $\kappa = 1$.

Applying a Suzuki–Trotter decomposition \cite{trotter1959product, 10.1143/PTP.56.1454} and a Hubbard–Stratonovich transformation \cite{Hubbard:1959ub} to the Hamiltonian, the model can be expressed in terms of an action $S$,
\begin{equation}
\label{eq:action}
S[\phi] = \sum_{x,t \in \Lambda} \frac{\phi_{xt}^2}{2 \delta U} - \log \det M[\phi] - \log \det M[-\phi] \,.
\end{equation}
Here, $\phi \in \mathbb{R}^\Lambda$ is a bosonic auxiliary field defined on the lattice $\Lambda = N_x \times N_t$, with physical sites $N_x$ and artificial sites $N_t$ arising from the Suzuki–Trotter decomposition. This decomposition introduces an error of $\mathcal{O}(\delta^2)$, where $\delta = \beta / N_t$ and $\beta$ is the inverse temperature.

Fermionic dynamics are encoded in the fermion matrix $M \in \mathbb{R}^{N_x N_t \times N_x N_t}$, whose elements are 
\begin{equation}
\label{eq:fermion_matrix}
M \left[ \phi \right]_{x’t’,xt} = \delta_{x’x}\delta_{t’t} - [e^{\delta h}]_{x’x} e^{\phi_{xt}} \mathcal{B}_{t’} \delta_{t’,t+1} \,,
\end{equation}
where $h$ is the hoping matrix, and $\mathcal{B}_t = 1$ for $1 < t \leq N_t$ and $\mathcal{B}_1 = -1$ implements anti-periodic temporal boundary conditions.

Owing to the sparsity of $M$, its determinant can be evaluated using an $LU$ decomposition,
\begin{equation}
\label{eq:lu}
\det M = \det L \cdot \det U \,. 
\end{equation}
As shown in Ref.~\cite{wynen2019}, the triangular structure of $L$ and $U$ allows an equivalent formulation:
\begin{equation}
\label{eq:sausage}
\det M = \det \left( \mathbbm{1} + A_1 A_2 \dots A_{N_t} \right) \,,
\end{equation}
where the product of matrices $A_t \in \mathbb{R}^{N_x \times N_x}$ is referred to as the \textit{sausage of matrices}. The individual matrices are given by
\begin{align}
\label{eq:Amatrices}
A_t = e^{\delta h} F_t \qquad \text{with} \qquad [F_t]_{xx’} = e^{\phi_{x,N_t-t}} \delta_{xx’} \,.
\end{align}

\section{Enhancing Numerical Stability with QR Decompositions}
\label{sec:qr}
\noindent
In~\cref{eq:Amatrices}, the matrices $A_t$ depend exponentially on the auxiliary field $\phi$ and the hopping matrix $h$. Consequently, the product $A \equiv A_1 A_2 \dots A_{N_t}$ develops a wide range of eigenvalues, resulting in a large condition number. While this behavior is mathematically well defined, it leads to numerical instabilities due to the accumulation of rounding errors during matrix multiplications. These instabilities propagate into the evaluation of the action and can produce unstable estimates of physical observables.

To mitigate this problem, we perform intermediate $QR$ decompositions~\cite{gander1980algorithms} during the computation of the matrix product. Here, $Q$ is an orthogonal matrix with unit condition number, while the triangular matrix $R$ contains the ill-conditioned part. This approach was previously employed in Ref.~\cite{determinant_bauer}, and we select the $QR$ decomposition over a singular value decomposition due to its significantly lower computational cost. 

Applying this procedure to the sausage of matrices, as shown in~\cref{alg:qr}, yields
\begin{equation}
QR = A \,.
\end{equation}
Rewriting the determinant from~\cref{eq:sausage} in terms of $Q$ and $R$,
\begin{equation}
\det \left( \mathbbm{1} + QR \right) = \det Q \cdot \det \left( Q^T + R \right) \,,
\end{equation}
enables a numerically stable evaluation. The same technique can also be used to compute the inverse of $\mathbbm{1} + QR$, which is required for several observables, including the gradient $\frac{\partial S}{\partial \phi_{xt}}$ used in HMC and in the training loop of the NF.
\begin{algorithm}
\caption{\label{alg:qr}Stable multiplication of a chain of matrices using $QR$ decomposition}
\begin{algorithmic}[1]
    \State $Q \gets \mathbbm{1}_{N_x \times N_x}$
    \State $R \gets \mathbbm{1}_{N_x \times N_x}$
    \For{each $A_t$ in $[ A_1, A_2, \dots, A_{N_t} ]$}
    \State $R \gets R \cdot A_t$
    \State $Q_\text{help}, R \gets$ QRDecomposition($R$)
    \State $Q \gets Q \cdot Q_\text{help}$
    \EndFor
    \end{algorithmic}
\end{algorithm}

\section{Methods}
\label{sec:method}

\subsection{Non-Equilibrium Markov Chain Monte Carlo}
\noindent
NE-MCMC is based on Jarzynski's equality~\cite{Jarzynski:1996oqb}, which relates the free-energy difference $\Delta F$ between an initial and a final macrostate of a statistical system undergoing an out-of-equilibrium evolution to an ensemble average,
\begin{equation}
\langle e^{-W} \rangle_\mathrm{f} = e^{-\Delta F} \,.
\end{equation}
Here, $W$ denotes the work performed on the system during the non-equilibrium process, and $\langle \cdot \rangle_\mathrm{f}$ indicates an average over all possible non-equilibrium trajectories.

Starting from an equilibrium Boltzmann distribution, a non-equilibrium Markov chain is constructed by updating the configuration using transition probabilities $P_{s(n)}$. The latter depends on a parameter $s$ of the action, which is varied at each step $n$ of the evolution. In particular, $P_{s(n)}(z_{n - 1} \to z_n)$, used to generate the configuration $z_n$ from $z_{n-1}$, satisfies detailed balance for the action $S_{s(n)}$ with the value of $s$ at the $n$-th step.
In this framework, the work $W$ is defined as the cumulative change of the action induced by variations of $s$ at fixed configurations,
\begin{equation}
\label{eq:jarzynski_work}
W = \sum_{n=0}^{\nstep-1} \left( S_{s(n+1)}\left[z_{n}\right] - S_{s(n)}\left[z_n\right] \right) \,,
\end{equation}
where $\nstep$ is the number of Monte Carlo steps along the evolution. 

It is important to notice that the system is never thermalized after the first step. Equilibirum expectation values are instead recovered via a reweighting formula,
\begin{equation}
\langle\mathcal{O}\rangle_p = \frac{\langle \mathcal{O} e^{-W} \rangle_\mathrm{f}}{\langle e^{-W} \rangle_\mathrm{f}},
\end{equation}
where $\langle\mathcal{O}\rangle_p$ denotes the expectation value with respect to the target distribution $p \propto e^{-S_{s(\nstep)}}$.

In the case of the Hubbard model, it is natural to define
\begin{equation}
    S_s[\phi] = \sum_{x,t \in \Lambda} \frac{\phi_{xt}^2}{2 \delta U} - s \left( \log \det M[\phi] + \log \det M[-\phi] \right) \,
\end{equation}
with $s\in [0,1]$. For $s=0$, the distribution is Gaussian, allowing trivially ergodic sampling, while the full action in~\cref{eq:action} is recovered at $s=1$. In the NE-MCMC simulations, we interpolate linearly from $s(0)=0$ to $s(\nstep)=1$.

\subsection{(Stochastic) Normalizing Flows}
\noindent
SNFs, introduced by Wu et al.~\cite{wu2020stochasticnormalizingflows}, combine deterministic NFs with stochastic NE-MCMC updates.
A NF $f_n: z_n \to f_n(z_n)$ induces a transformation of probability densities according to
\begin{equation}
\label{eq:nf_density}
\log q_{n+1}(f_n(z_n)) = \log q_n(z_n) - \log \det \left| \frac{\partial f_n(z_n)}{\partial z} \right| \,,
\end{equation}
where the NF allows an easily tractable Jacobian determinant that encodes the change in volume. A SNF is constructed by alternating blocks of flow transformations and NE-MCMC updates. The resulting chain can be represented as
\begin{equation}
\label{eq:snf_chain}
z_0 \overset{\text{NF}}{\longrightarrow} f_0(z_0) \overset{\text{MC}}{\longrightarrow} z_1 \overset{\text{NF}}{\longrightarrow} f_1(z_1) \overset{\text{MC}}{\longrightarrow} z_2 \overset{\text{NF}}{\longrightarrow} \dots \overset{\text{MC}}{\longrightarrow} z_{\nstep} \equiv \phi \,.
\end{equation}
The output density $q_{\nstep}$ is obtained by combining the change-of-variables formula in~\cref{eq:nf_density} with the accumulated work in~\cref{eq:jarzynski_work},
\begin{equation}
\log q_{\nstep}(\phi) = \log q_0(z_0) - W - \sum_{n=0}^{\nstep - 1} \log \det \left| \frac{\partial f_n(z_n)}{\partial z} \right|  \,.
\end{equation}
Training of the SNF is performed by minimizing the reverse Kullback–Leibler divergence \cite{kullback1951information}, given by
\begin{equation}
\text{KL}( q_{\nstep} || p) = \langle S_{s(\nstep)}[\phi] + \log q_{\nstep}(\phi) \rangle_{\phi \sim q_{\nstep}} \,.
\end{equation}

\section{Results}
\label{sec:results}
\noindent
In this section, we present our results on the numerical stability of the action and the scalability of NFs, SNFs and NE-MCMC.

\subsection{Computational Cost of the Action}
\label{sec:action_cost}
\begin{figure}[t]
    \centering
    \begin{subfigure}[b]{0.49\textwidth}
    \centering
    \includegraphics[width=\textwidth]{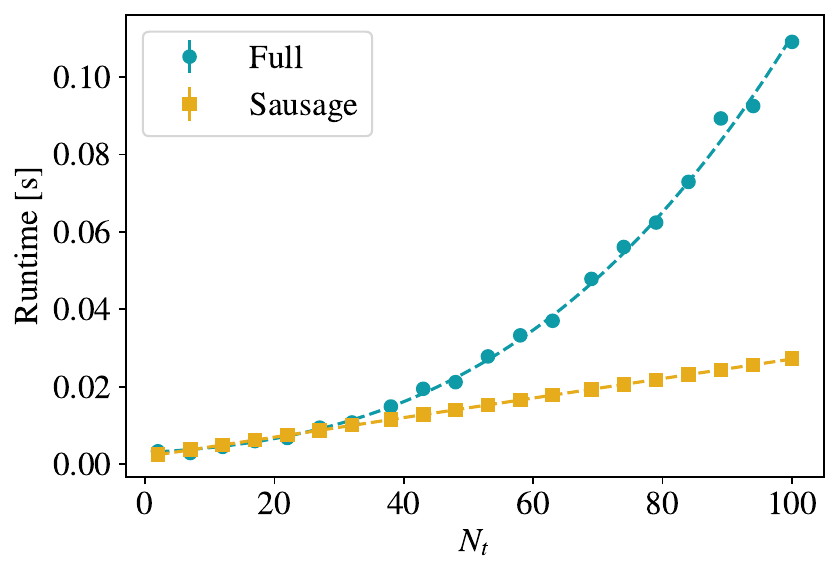}
    \caption{Computational cost}
    \end{subfigure}
    \hfill
    \begin{subfigure}[b]{0.49\textwidth}
    \centering
    \includegraphics[width=\textwidth]{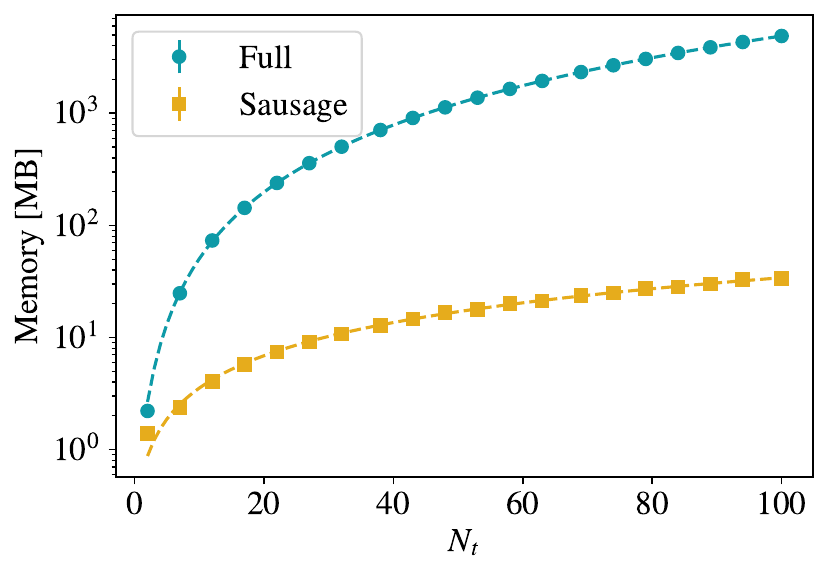}
    \caption{Memory consumption}
    \end{subfigure}
    \caption{Scaling of computational cost (a) and memory consumption (b) for the full fermion matrix (blue) and the sausage formalism (orange) for different $N_t$ at fixed $N_x = 4$. Dashed lines correspond to fits based on the scaling behavior from \cref{tab:action_scaling}. Error bars are smaller than the marker size.}
    \label{fig:action_scaling_nt}
\end{figure}

\noindent
As discussed in~\cref{sec:hubbard}, the determinant of the fermion matrix can be expressed using the sausage of matrices. This formulation exploits the sparsity of the full fermion matrix, significantly reducing both computational cost and memory consumption. A scaling analysis of these quantities is shown in~\cref{fig:action_scaling_nt}, while the key scaling behavior is summarized in~\cref{tab:action_scaling}. Using the sausage formalism reduces the computational cost by a factor of $N_t^2$ and memory usage by a factor of $N_t$.

\begin{table}[t]
    \centering
    \begin{tabular}{c|cc}
    & Full & Sausage \\
    \hline
    Computational cost & $N_x^3 N_t^3$ & $N_x^3 N_t$ \\
    Memory consumption & $N_x^2 N_t^2$ & $N_x^2 N_t$
    \end{tabular}
    \caption{Scaling behavior of computational cost and memory consumption for the full fermion matrix and the sausage formalism. Both quantities improve in terms of $N_t$.}
    \label{tab:action_scaling}
\end{table}
\begin{figure}[t]
    \centering
    \includegraphics[width=0.5\linewidth]{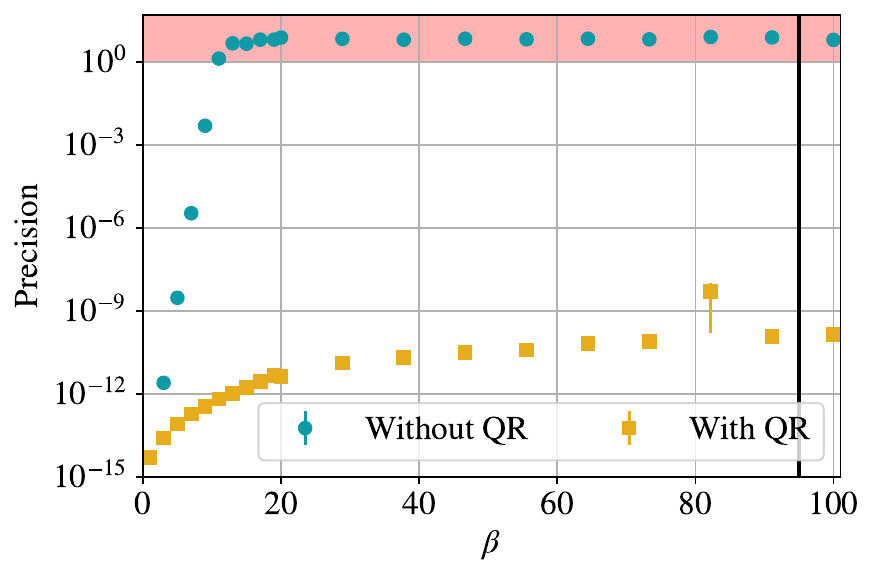}
    \caption{Precision of the Hubbard action in the sausage formalism with (orange) and without (blue) QR decomposition for different $\beta$. The black line marks graphene at room temperature, and the red shaded area indicates impracticable precision. The system is evaluated at $U=2$, $N_x = 20$, and $\delta = 0.5$. For most data points, the error bars are smaller than the markers.}
    \label{fig:stable_action}
\end{figure}

\subsection{Numerical Stability of the Action at Low Temperatures}
\label{sec:results_stability}
\noindent
The numerical instability of the action in the sausage formalism introduces a bias that becomes more severe at larger inverse temperatures $\beta$ and larger numbers of temporal sites $N_t$. To quantify the numerical precision, we define
\begin{equation}
\label{eq:precision}
\text{Precision} \equiv \sqrt{\frac{1}{N_\text{cfg}} \sum_{i=1}^{N_\text{cfg}} \left( S\left[\phi^{(i)}\right] - S\left[\phi^{(i)}_r\right] \right)^2},
\end{equation}
which measures the root-mean-square difference between the action computed for a configuration $\phi$ and a time-shifted configuration $\phi_r$, with $(\phi_r)_{x,t} \equiv \phi_{x,t+1}$. In exact computations, where instabilities are absent, this quantity would be zero due to time-translational symmetry.

\cref{fig:stable_action} shows the numerical precision with and without $QR$ decompositions as a function of $\beta$, while keeping $\delta$ fixed (thus increasing $N_t$). Without QR stabilization, the action becomes inaccurate for $\beta > 10$, whereas the QR decomposition maintains good precision up to $\beta = 100$. The vertical black line indicates graphene at room temperature, serving as a target for future simulations.

\subsection{Scaling of the Training: (S)NFs}
\label{sec:scaling_snf}
\begin{figure}[t]
    \centering
    \begin{subfigure}[b]{0.493\textwidth}
    \centering
    \includegraphics[width=\textwidth]{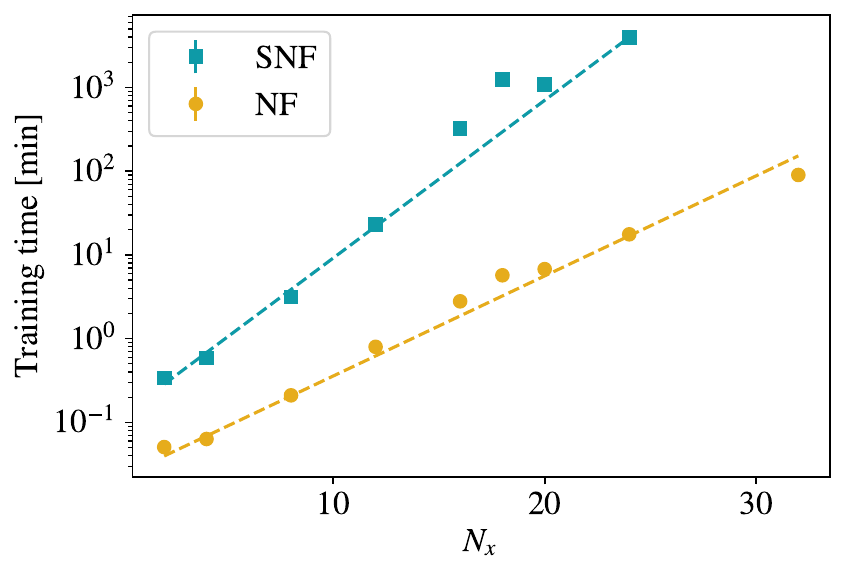}
    \caption{NF and SNF at $\beta=5$, $N_t = 20$}
    \end{subfigure}
    \hfill
    \raisebox{-0.3mm}{
    \begin{subfigure}[b]{0.48\textwidth}
        \centering
        \includegraphics[width=\textwidth]{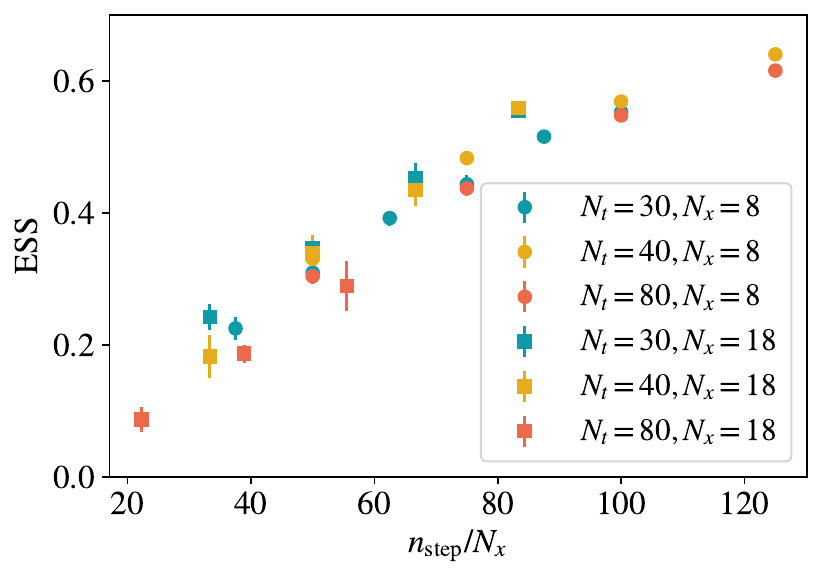}
        \caption{NE-MCMC at $\beta=12$}
    \end{subfigure}
    }
    \caption{Scaling of the training cost of (S)NFs for a fixed architecture (left) and scaling of the sampling cost of NE-MCMC (right) at $U=2$.}
    \label{fig:scaling_snf}
\end{figure}

\noindent
Here, we quantify the scaling of (S)NFs by measuring the computation cost, i.e., the training time, required to reach a given measure of quality. In this setup, we report the training cost until the sampler achieves a 20\,\% acceptance rate in an accept-reject step. In~\cref{fig:scaling_snf}(a), we compare RealNVP \cite{dinh2017densityestimationusingreal}, a standard NF architecture, with an SNF having a fixed $\nstep=3$ at $U=2$, $\beta=5$, and $N_t=20$. Both flows exhibit an exponential scaling with system size; however, the SNF shows an additional offset due to the extra computations of the action during the training loop.

Importantly, the present study is performed for a fixed architecture. Future work will explore how the training cost scales with the number of Monte Carlo steps and the number of parameters in the NF layers.

\subsection{Scaling of the sampling: NE-MCMC}
\label{sec:scaling_jarzynski}
\noindent
It has been established in various physical setups that NE-MCMC exhibits an approximately linear scaling with the lattice volume~\cite{Bulgarelli_2025, Bonanno:2025pdp}. More precisely, to maintain a given sampling quality (here quantified by the effective sampling size ($\mathrm{ESS}$)) as the volume increases, the number of steps in the non-equilibrium evolution must grow linearly with the system size. 

This scaling behavior is illustrated in~\cref{fig:scaling_snf}(b), which shows the $\mathrm{ESS}$ of NE-MCMC simulations at $U=2$, $\beta=12$, and various lattice sizes $N_x \times N_t$. The data collapse indicates that the $\mathrm{ESS}$ is approximately a function of $\nstep / N_x$, confirming the expected linear scaling with the spatial extent. Interestingly, there is no significant dependence on $N_t$. This is not in conflict with previous observations~\cite{Bulgarelli_2025, Bonanno:2025pdp}, as space and time play distinct roles in the Hubbard model. 

This scaling behavior is expected to carry over to SNFs, while the NF transformation reduces the computational cost. Consequently, SNFs offer a promising approach for simulating the Hubbard model on larger lattices, despite the increased training cost at fixed step size.

\section{Conclusion}
\label{sec:conclusion}
\noindent
In this work, we have studied the numerical stability and scalability of NFs, SNFs, and NE-MCMC methods for the Hubbard model. An important step toward improved scalability is the \textit{sausage} formalism, which enables linear scaling in the temporal lattice extent $N_t$ when computing the fermion determinant. Numerical stability, even at low temperatures, can be ensured by performing QR decompositions between matrix multiplications. 

Our numerical results show that NFs and SNFs, when operated with a fixed number of Monte Carlo steps, exhibit an exponential scaling of training time with the spatial lattice extent $N_x$. In contrast, NE-MCMC shows the expected linear scaling with the lattice volume and therefore currently provides the most scalable and robust sampling strategy for the Hubbard model.

For future work, we plan to conduct a detailed analysis of the scaling behavior of SNFs when the number of Monte Carlo steps is allowed to increase with $N_x$. While the present results indicate unfavorable scaling at fixed Monte Carlo depth, we expect that, once this restriction is lifted, both training and sampling costs of SNFs can achieve linear scaling with the lattice volume, similar to NE-MCMC. Under these conditions, SNFs have the potential to become a competitive and fully scalable tool for ergodic sampling of the Hubbard model.

\acknowledgments
\noindent
The authors thank Kim A. Nicoli for discussions during the initial phase of this project. The authors gratefully acknowledge the access to the Marvin cluster of the University of Bonn. This project was supported by the Deutsche Forschungsgemeinschaft (DFG, German Research Foundation) as part of the CRC 1639 NuMeriQS -- project no.\ 511713970. LV acknowledges support from the SFT
Scientific Initiative of INFN and the Simons Foundation grant 994300.

\bibliographystyle{JHEP}
\bibliography{refs}

@article{wynen2019,
    author = {Wynen, Jan-Lukas and Berkowitz, Evan and K{\"o}rber, Christopher and L{\"a}hde, Timo A. and Luu, Thomas},
    title = "{Avoiding Ergodicity Problems in Lattice Discretizations of the Hubbard Model}",
    eprint = "1812.09268",
    archivePrefix = "arXiv",
    primaryClass = "cond-mat.str-el",
    doi = "10.1103/PhysRevB.100.075141",
    journal = "Phys. Rev. B",
    volume = "100",
    number = "7",
    pages = "075141",
    year = "2019"
}

@article{hubbard,
author = {Hubbard, J.},
title = {{Electron correlations in narrow energy bands}},
journal = {Proc. Roy. Soc. A},
volume = {276},
number = {1365},
pages = {238-257},
year = {1963},
doi = {10.1098/rspa.1963.0204},
}

@article{Arovas_2022,
   title={{The Hubbard Model}},
   volume={13},
   ISSN={1947-5462},
   url={http://dx.doi.org/10.1146/annurev-conmatphys-031620-102024},
   DOI={10.1146/annurev-conmatphys-031620-102024},
   number={1},
   journal={Annu. Rev. Condens. Matter Phys.},
   publisher={Annual Reviews},
   author={Arovas, Daniel P. and Berg, Erez and Kivelson, Steven A. and Raghu, Srinivas},
   year={2022},
   month=mar, pages={239--274},
eprint="2103.12097"
}

@article{PhysRevD.100.034515,
  title = {{Flow-based generative models for Markov chain Monte Carlo in lattice field theory}},
  author = {Albergo, M. S. and Kanwar, G. and Shanahan, P. E.},
  journal = {Phys. Rev. D},
  volume = {100},
  issue = {3},
  pages = {034515},
  numpages = {13},
  year = {2019},
  month = {Aug},
  publisher = {American Physical Society},
  doi = {10.1103/PhysRevD.100.034515},
  url = {https://link.aps.org/doi/10.1103/PhysRevD.100.034515},
eprint="1904.12072",
}

@article{PhysRevE.101.023304,
  title = {{Asymptotically unbiased estimation of physical observables with neural samplers}},
  author = {Nicoli, Kim A. and Nakajima, Shinichi and Strodthoff, Nils and Samek, Wojciech and M\"uller, Klaus-Robert and Kessel, Pan},
  journal = {Phys. Rev. E},
  volume = {101},
  issue = {2},
  pages = {023304},
  numpages = {10},
  year = {2020},
  month = {Feb},
  publisher = {American Physical Society},
  doi = {10.1103/PhysRevE.101.023304},
  url = {https://link.aps.org/doi/10.1103/PhysRevE.101.023304},
eprint="1910.13496"
}

@article{bialas2024r,
  title = {{R\'enyi entanglement entropy of a spin chain with generative neural networks}},
  author = {Bia\l{}as, Piotr and Korcyl, Piotr and Stebel, Tomasz and Zapolski, Dawid},
  journal = {Phys. Rev. E},
  volume = {110},
  issue = {4},
  pages = {044116},
  numpages = {9},
  year = {2024},
  month = {Oct},
  publisher = {American Physical Society},
  doi = {10.1103/PhysRevE.110.044116},
  url = {https://link.aps.org/doi/10.1103/PhysRevE.110.044116},
eprint = "2406.06193",
}

@article{
doi:10.1126/science.aaw1147,
author = {Frank Noé  and Simon Olsson  and Jonas Köhler  and Hao Wu },
title = {{Boltzmann generators: Sampling equilibrium states of many-body systems with deep learning}},
journal = {Science},
volume = {365},
number = {6457},
pages = {eaaw1147},
year = {2019},
doi = {10.1126/science.aaw1147},
eprint = {1812.01729}
}

@article{kobyzev2020normalizing,
  author={Kobyzev, Ivan and Prince, Simon J.D. and Brubaker, Marcus A.},
  journal={IEEE Transactions on Pattern Analysis and Machine Intelligence}, 
  title={{Normalizing Flows: An Introduction and Review of Current Methods}}, 
  year={2021},
  volume={43},
  number={11},
  pages={3964--3979},
  doi={10.1109/TPAMI.2020.2992934},
eprint={1908.09257}
}

@inproceedings{rezende2015variational,
	title        = {{Variational Inference with Normalizing Flows}},
	author       = {Rezende, Danilo and Mohamed, Shakir},
	year         = 2015,
	booktitle    = {Proceedings of the 32nd International Conference on Machine Learning},
	publisher    = {PMLR},
	volume       = 37,
	pages        = {1530--1538},
eprint = "1505.05770"
}

@InProceedings{oord2016pixelrecurrentneuralnetworks,
  title = 	 {{Pixel Recurrent Neural Networks}},
  author = 	 {van den Oord, Aäron and Kalchbrenner, Nal and Kavukcuoglu, Koray},
  booktitle = 	 {Proceedings of The 33rd International Conference on Machine Learning},
  pages = 	 {1747--1756},
  year = 	 {2016},
  volume = 	 {48},
  series = 	 {Proceedings of Machine Learning Research},
  month = 	 {20--22 Jun},
  publisher =    {PMLR},
eprint="1601.06759"
}

@article{Hubbard:1959ub,
  title = {{Calculation of Partition Functions}},
  author = {Hubbard, J.},
  journal = {Phys. Rev. Lett.},
  volume = {3},
  issue = {2},
  pages = {77--78},
  numpages = {0},
  year = {1959},
  month = {Jul},
  publisher = {American Physical Society},
  doi = {10.1103/PhysRevLett.3.77},
  url = {https://link.aps.org/doi/10.1103/PhysRevLett.3.77}
}

@article{PhysRevLett.126.032001,
  title = {{Estimation of Thermodynamic Observables in Lattice Field Theories with Deep Generative Models}},
  author = {Nicoli, Kim A. and Anders, Christopher J. and Funcke, Lena and Hartung, Tobias and Jansen, Karl and Kessel, Pan and Nakajima, Shinichi and Stornati, Paolo},
  journal = {Phys. Rev. Lett.},
  volume = {126},
  issue = {3},
  pages = {032001},
  numpages = {6},
  year = {2021},
  month = {Jan},
  publisher = {American Physical Society},
  doi = {10.1103/PhysRevLett.126.032001},
  url = {https://link.aps.org/doi/10.1103/PhysRevLett.126.032001},
eprint="2007.07115"
}

@article{PhysRevLett.122.080602,
  title = {{Solving Statistical Mechanics Using Variational Autoregressive Networks}},
  author = {Wu, Dian and Wang, Lei and Zhang, Pan},
  journal = {Phys. Rev. Lett.},
  volume = {122},
  issue = {8},
  pages = {080602},
  numpages = {6},
  year = {2019},
  month = {Feb},
  publisher = {American Physical Society},
  doi = {10.1103/PhysRevLett.122.080602},
  url = {https://link.aps.org/doi/10.1103/PhysRevLett.122.080602},
eprint="1809.10606"
}

@article{cranmer2023advances,
  title={{Advances in machine-learning-based sampling motivated by lattice quantum chromodynamics}},
  author={Cranmer, Kyle and Kanwar, Gurtej and Racani{\`e}re, S{\'e}bastien and Rezende, Danilo J and Shanahan, Phiala E},
  journal={Nature Reviews Physics},
  volume={5},
  number={9},
  pages={526--535},
  year={2023},
  publisher={Nature Publishing Group UK London},
eprint="2309.01156",
doi="10.1038/s42254-023-00616-w"
}

@inproceedings{gebauer1,
 author = {Gebauer, Niklas and Gastegger, Michael and Sch\"{u}tt, Kristof},
 booktitle = {Advances in Neural Information Processing Systems},
 title = {{Symmetry-adapted generation of 3d point sets for the targeted discovery of molecules}},
 volume = {32},
 year = {2019},
eprint="1906.00957"
}

@article{gebauer3,
	author = {Gebauer, Niklas W. A. and Gastegger, Michael and Hessmann, Stefaan S. P. and M{\"u}ller, Klaus-Robert and Sch{\"u}tt, Kristof T.},
	date = {2022/02/21},
	doi = {10.1038/s41467-022-28526-y},
	id = {Gebauer2022},
	isbn = {2041-1723},
	journal = {Nature Commun.},
	number = {1},
	pages = {973},
	title = {{Inverse design of 3d molecular structures with conditional generative neural networks}},
	url = {https://doi.org/10.1038/s41467-022-28526-y},
	volume = {13},
	year = {2022},
    eprint = "2109.04824",
    archivePrefix = "arXiv",
    primaryClass = "cs.LG",
}

@article{Bulgarelli:2024yrz,
    author = {Bulgarelli, Andrea and Cellini, Elia and Jansen, Karl and K{\"u}hn, Stefan and Nada, Alessandro and Nakajima, Shinichi and Nicoli, Kim A. and Panero, Marco},
    title = "{Flow-Based Sampling for Entanglement Entropy and the Machine Learning of Defects}",
    eprint = "2410.14466",
    archivePrefix = "arXiv",
    primaryClass = "quant-ph",
    doi = "10.1103/PhysRevLett.134.151601",
    journal = "Phys. Rev. Lett.",
    volume = "134",
    number = "15",
    pages = "151601",
    year = "2025"
}

@article{Caselle:2023mvh,
    author = "Caselle, Michele and Cellini, Elia and Nada, Alessandro",
    title = "{Sampling the lattice Nambu-Goto string using Continuous Normalizing Flows}",
    eprint = "2307.01107",
    archivePrefix = "arXiv",
    primaryClass = "hep-lat",
    doi = "10.1007/JHEP02(2024)048",
    journal = "JHEP",
    volume = "02",
    pages = "048",
    year = "2024"
}

@article{Caselle:2022acb,
    author = "Caselle, Michele and Cellini, Elia and Nada, Alessandro and Panero, Marco",
    title = {{Stochastic normalizing flows as non-equilibrium transformations}},
    eprint = "2201.08862",
    archivePrefix = "arXiv",
    primaryClass = "hep-lat",
    doi = "10.1007/JHEP07(2022)015",
    journal = "JHEP",
    volume = "07",
    pages = "015",
    year = "2022"
}

@article{Caselle:2024ent,
    author = "Caselle, Michele and Cellini, Elia and Nada, Alessandro",
    title = "{Numerical determination of the width and shape of the effective string using Stochastic Normalizing Flows}",
    eprint = "2409.15937",
    archivePrefix = "arXiv",
    primaryClass = "hep-lat",
    doi = "10.1007/JHEP02(2025)090",
    journal = "JHEP",
    volume = "02",
    pages = "090",
    year = "2025"
}

@article{trotter1959product,
  title={{On the product of semi-groups of operators}},
  author={Trotter, Hale F},
  journal={Proc. Amer. Math. Soc.},
  volume={10},
  number={4},
  pages={545--551},
  year={1959},
  publisher={JSTOR},
doi={10.2307/2033649}
}

@article{10.1143/PTP.56.1454,
    author = {Suzuki, Masuo},
    title = {{Relationship between d-Dimensional Quantal Spin Systems and (d+1)-Dimensional Ising Systems: Equivalence, Critical Exponents and Systematic Approximants of the Partition Function and Spin Correlations}},
    journal = {Prog. Theor. Phys.},
    volume = {56},
    number = {5},
    pages = {1454--1469},
    year = {1976},
    month = {11},
    issn = {0033-068X},
    doi = {10.1143/PTP.56.1454},
    url = {https://doi.org/10.1143/PTP.56.1454},
}

@article{Schuh:2025gks,
    author = "Schuh, Dominic and Kreit, Janik and Berkowitz, Evan and Funcke, Lena and Luu, Thomas and Nicoli, Kim A. and Rodekamp, Marcel",
    title = "{Simulating Correlated Electrons with Symmetry-Enforced Normalizing Flows}",
    eprint = "2506.17015",
    archivePrefix = "arXiv",
    primaryClass = "cond-mat.str-el",
    month = "6",
    year = "2025"
}

@article{Kreit:2025cos,
    author = "Kreit, Janik and Schuh, Dominic and Nicoli, Kim A. and Funcke, Lena",
    title = "{SESaMo: Symmetry-Enforcing Stochastic Modulation for Normalizing Flows}",
    eprint = "2505.19619",
    archivePrefix = "arXiv",
    primaryClass = "cs.LG",
    month = "5",
    year = "2025"
}

@inproceedings{wu2020stochasticnormalizingflows,
 author = {Wu, Hao and K\"{o}hler, Jonas and Noe, Frank},
 booktitle = {Advances in Neural Information Processing Systems},
 pages = {5933--5944},
 title = "{Stochastic Normalizing Flows}",
 eprint = "2002.06707",
 archivePrefix = "arXiv",
 primaryClass = "stat.ML",
 volume = {33},
 year = {2020}
}

@article{Bulgarelli_2025,
   author = "Bulgarelli, Andrea and Cellini, Elia and Nada, Alessandro",
    title = "{Scaling of stochastic normalizing flows in SU(3) lattice gauge theory}",
    eprint = "2412.00200",
    archivePrefix = "arXiv",
    primaryClass = "hep-lat",
    doi = "10.1103/PhysRevD.111.074517",
    journal = "Phys. Rev. D",
    volume = "111",
    number = "7",
    pages = "074517",
    year = "2025"
}

@article{gander1980algorithms,
  title={Algorithms for the QR decomposition},
  author={Gander, Walter},
  journal={Res. Rep},
  volume={80},
  number={02},
  pages={1251--1268},
  year={1980}
}

@article{Bonanno:2024udh,
    author = "Bonanno, Claudio and Nada, Alessandro and Vadacchino, Davide",
    title = "{Mitigating topological freezing using out-of-equilibrium simulations}",
    eprint = "2402.06561",
    archivePrefix = "arXiv",
    primaryClass = "hep-lat",
    doi = "10.1007/JHEP04(2024)126",
    journal = "JHEP",
    volume = "04",
    pages = "126",
    year = "2024"
}

@article{Caselle:2016wsw,
    author = "Caselle, Michele and Costagliola, Gianluca and Nada, Alessandro and Panero, Marco and Toniato, Arianna",
    title = "{Jarzynski{\textquoteright}s theorem for lattice gauge theory}",
    eprint = "1604.05544",
    archivePrefix = "arXiv",
    primaryClass = "hep-lat",
    reportNumber = "CP3-ORIGINS-2016-020, DIAS-2016-20",
    doi = "10.1103/PhysRevD.94.034503",
    journal = "Phys. Rev. D",
    volume = "94",
    number = "3",
    pages = "034503",
    year = "2016"
}

@article{Jarzynski:1996oqb,
      author         = "Jarzynski, Christopher",
      title          = "{Nonequilibrium Equality for Free Energy Differences}",
      journal        = "Phys. Rev. Lett.",
      volume         = "78",
      issue          = "14",
      pages          = "2690--2693",
      year           = "1997",
      publisher      = "American Physical Society",
      doi            = "10.1103/PhysRevLett.78.2690",
      url            = "http://link.aps.org/doi/10.1103/PhysRevLett.78.2690",
      eprint         = "cond-mat/9610209",
      archivePrefix  = "arXiv",
      primaryClass   = "cond-mat",
}

@article{
noneq_jerome,
author = {Jerome P. Nilmeier  and Gavin E. Crooks  and David D. L. Minh  and John D. Chodera },
title = {Nonequilibrium candidate Monte Carlo is an efficient tool for equilibrium simulation},
journal = {Proceedings of the National Academy of Sciences},
volume = {108},
number = {45},
pages = {E1009-E1018},
year = {2011},
doi = {10.1073/pnas.1106094108},
eprint = {1105.2278}
}

@article{determinant_bauer,
	title={Fast and stable determinant quantum Monte Carlo},
	author={Carsten Bauer},
	journal={SciPost Phys. Core},
	volume={2},
	pages={011},
	year={2020},
	publisher={SciPost},
	doi={10.21468/SciPostPhysCore.2.2.011},
	url={https://scipost.org/10.21468/SciPostPhysCore.2.2.011},
    eprint = "2003.05286",
}

@article{kullback1951information,
  title={On information and sufficiency},
  author={Kullback, Solomon and Leibler, Richard A},
  journal={The annals of mathematical statistics},
  volume={22},
  number={1},
  pages={79--86},
  year={1951},
  publisher={JSTOR}
}

@article{
dinh2017densityestimationusingreal,
title={Density estimation using {R}eal {NVP}},
author={Laurent Dinh and Jascha Sohl-Dickstein and Samy Bengio},
year={2017},
journal={ICLR},
eprint = "1605.08803",
archivePrefix = "arXiv",
primaryClass = "cs.LG",
}

@article{Bonanno:2025pdp,
    author = "Bonanno, Claudio and Bulgarelli, Andrea and Cellini, Elia and Nada, Alessandro and Panfalone, Dario and Vadacchino, Davide and Verzichelli, Lorenzo",
    title = "{Scaling flow-based approaches for topology sampling in $\mathrm{SU}(3)$ gauge theory}",
    eprint = "2510.25704",
    archivePrefix = "arXiv",
    primaryClass = "hep-lat",
    month = "10",
    year = "2025"
}

\end{document}